\documentclass[aps,nofootinbib,onecolumn,citeautoscript]{revtex4-2}
\usepackage{amsmath,amssymb}
\usepackage{enumerate}
\usepackage{xcolor}
\usepackage[colorlinks=true,citecolor=red,linkcolor=blue,linkbordercolor=green]{hyperref}
\usepackage{breqn}
\usepackage{setspace}
\usepackage{enumerate}
\usepackage{graphicx}
\usepackage{float}

\date{\today}

\begin{document}
\title{Efficacy of Moriya interaction to free the bound entangled state}
\author{Kapil K. Sharma$^\ast$, Suprabhat Sinha$^\dagger$ and Krishna Chandra$^\#$ \\\vspace{0.4cm}
\textit{$^\ast$School of Computer Science, Engineering and Applications, \\
DY Patil International University, Akurdi, \\
 Pune-411044, India} \\
E-mail: iitbkapil@gmail.com \\\vspace{0.4cm}
\textit{$^\dagger$Department of Electronics, \\
West Bengal State University, \\
Barasat, Kolkata-700126, India} \\ 
Email: suprabhatsinha64@gmail.com \\\vspace{0.4cm}
\textit{$^\#$Department of Physics,\\
Government Arts Science Commerce College Nagda,\\ Ujjain-456335, India} \\
Email: 1256krishna@gmail.com}

\begin{abstract}
The current work shows the efficacy of Dzyaloshinshkii-Moriya (DM) interaction to free the bound entanglement. Based on the work [\textit{Sharma, K.K., Pandey, S.N., Quant. Info. Proc. \textbf{15}, 1539 (2016)}], we present further results in two qutrits bound entangled state proposed by Jurkovaski \textit{et al.} We consider a closed system of two qutrits and an auxiliary qutrit which interacts with either one of the two qutrits in a closed system. We erase the auxiliary qutrit from the system by doing partial trace operation and open system dynamics has been studied. We have found, the probability amplitude of auxiliary qutrit does not affect the system, while DM interaction plays a major role to govern the dynamics. The realignment and CCNR criteria have been used to detect the bound entanglement in the state, while for quantification of entanglement the negativity has been used. We explored the dynamics with all the possible cases of Jurkowski \textit{et al.} bound entangled state. 
\end{abstract}
\maketitle

\section{Introduction}
Entanglement is one of the most fundamental tools for future quantum technologies \cite{EPR1935,Neilsen2000,ml}. Many applications of entanglement like quantum teleportation, quantum cryptography, quantum games, quantum imaging, etc. have been investigated  \cite{CBennet1933,AkEkert1991,qs,qi}. All these applications need maximally entangled states for the perfect execution of quantum information. But quantum states are too evasive such that they may lose their entanglement when interacting with the external environment. So dynamics of entanglement under various environmental interactions in different quantum states play an important role to process the quantum information. In this paper, we treat DM interaction as an environmental interaction. For the dimension $\geq 3$, the quantum states have been divided into two categories such as free entangled and bound entangled state \cite{Horodecki_1,Horodecki_2}. Free entangled states are distillable states and can be effectively utilized for quantum information processing \cite{Dist}. It is known that all the quantum states in $2 \otimes 2$ and $2 \otimes 3$ dimensions are distillable. While on the other hand, the bound entangled states are noisy states and hard to distill. So it is difficult to use these states for practical quantum information processing. However these states increase the teleportation fidelity \cite{Activation_1,Activation_2,Activation_3}. Recently, the bound entanglement have been realized experimentally in quantum information processing \cite{EB_0,EB_1,EB_2}. At present, there are many mathematical tools available to characterize and detect the entanglement for bipartite systems in $2\otimes 2$ and $2\otimes 3$ dimensions. Negativity is one of these and is a good measure of entanglement quantification \cite{A.Peres,schidmt,Perse1996,negativity,locc}. But for the dimensions $\geq 3$, the characterization and detection of entanglement is still an open problem. Various criteria have been given to detect the bound entanglement such as realignment criteria, computable cross norm realignment criteria (CCNR), etc. for higher dimensions $(\geq 3)$ \cite{PPT,range_citeria,Topical,Realign,Cross_norm,Cross_norm_update}.

The entanglement dynamics of two-qubit pair under DM interaction \cite{DMinteraction,DMmoriya,Tmoriya2_1960}, by taking a third controller qubit, has been studied by Zheng Qiang \textit{et al.} \cite{ZQiang1,ZQiang2,ZQiang3}. The third controller qubit interacts with the qubit of the pair through DM interaction.
They have shown that by adjusting the state (i.e. probability amplitude) of either controller qubit or qutrit and DM interaction strength, one can control the entanglement between two-qubit pair. This kind of study is not only useful in qubits but also useful in the quantum systems with the dimensions $(d\otimes d)$, $d\geq 3$ and in hybrid quantum systems with arbitrary dimensions. Recently we have studied the dynamics in the hybrid qubit-qutrit system under DM interaction \cite{k1,k2,k3,k4,k5}. We have found that the state (i.e. probability amplitude) of auxiliary qubit does not play any role in entanglement dynamics in some cases. So in this case one can avoid the intention to prepare the specific state of the auxiliary qubit to manipulate the entanglement in the system. Further, we also have shown the efficacy of DM interaction to free the bound entanglement in the context of Horodecki et al. bound entangled states \cite{k6}. Here we recall that bound entangled states take place in Hilbert space with the dimension $(\geq 3)$. The dimension $(d=3)$ corresponds to qutrit. It has been proved that the usage of qutrit is more secure against a symmetric attack on a quantum key distribution protocol\cite{qut}. A quantum communication complexity protocol is also proposed by using two entangled qutrits \cite{pro}. So a qutrit system is of special interest as it has great manifestation and best fit into the dimensionality of Hilbert space and increases the computing power. With the dimensions $(d\otimes d)$, $d\geq 3$, we have bound entanglement in nature and its dynamics under various circumstances is the subject of investigation. The dynamics of bound entangled states has been studied and the phenomenon of distillability sudden death has also been observed \cite{b2,b5,b7,b3,b6,b4,b1,sq1}. 

Motivated from the above mentioned studies, we study the dynamics of two qutrits bound entangled state under DM interaction by taking an auxiliary qutrit that interacts with any one of the qutrits. The auxiliary qutrit assists to establish the environmental DM interaction.  Here we consider the bound entangled state investigated by Jurkowski \textit{et al.} \cite{Qutrit_states}. To the best of our knowledge, the study of distillability of Jurkowski \textit{et al.} bound entangled state through DM interaction has not been reported as yet. The main goal of the present study is to show that the external DM interaction can be used as a useful resource to produce the free entanglement in two qutrits bound entangled state during the dynamic evolution. Once the states are free then these can be easily distilled. We also find that the state (probability amplitude) of auxiliary qutrit does not play any role in entanglement dynamics in two qutrits bound entangled state. The present study can be useful in investigating the free entanglement extraction from Bound entangled state by DM interaction.

The plan of the paper is as follows. In Sect. 2 we present the two qutrits bound entangled state provided by Jurkowski \textit{et al.} and its unitary dynamics with the Hamiltonian of the system. Negativity, realignment criteria, and the computable cross-norm and realignment (CCNR) criterion are discussed in Sect. 3. In Sect. 4 open system dynamics of the system is presented with the significant results. Finally, in the last section, we have discussed the conclusion of our current work.

\section{Bound entangled state and unitary dynamics}
In this section, we discuss the bound entangled state and its interaction with auxiliary qutrit which we use in the current work. The state is a qutrit(A)-qutrit(B) bound entangled state investigated by Jurkowski \textit{et al.} \cite{Qutrit_states}. The density matrix of the state can be written as,

\begin{equation}
\rho(\epsilon_{1},\epsilon_{2},\epsilon_{3})=\frac{1}{N}\left[ \begin{array}{ccccccccc}
1 &0&0&0&1&0&0&0&1 \\
0&\epsilon_{1}&0&0&0&0&0&0&0 \\
0&0&\epsilon_{3}^{-1}&0&0&0&0&0&0 \\
0&0&0&\epsilon_{1}^{-1}&0&0&0&0&0 \\
1&0&0&0&1&0&0&0&1 \\
0&0&0&0&0&\epsilon_{2}&0&0&0 \\
0&0&0&0&0&0&\epsilon_{3}&0&0 \\
0&0&0&0&0&0&0&\epsilon_{2}^{-1}&0 \\
1&0&0&0&1&0&0&0&1 \\
\end{array}
\right ]. \ \ \ \label{bes}
\end{equation} 
Where $N=(1+\epsilon_{1}+\epsilon_{3}^{-1}+\epsilon_{1}^{-1}+1+\epsilon_{2}+\epsilon_{3}+\epsilon_{2}^{-1}+1)$ is the normalization constant and the state satisfy the trace condition given below,
\begin{equation}
Tr[\rho]=1. \label{tr1}
\end{equation}
The state depends on the three parameters $\epsilon_{1}$, $\epsilon_{2}$, $\epsilon_{3}$. For the condition $\epsilon_{1}=\epsilon_{2}=\epsilon_{3}=1$, the state is separable. 

In the present work, we consider an auxiliary qutrit which interacts through DM interaction with any qutrit of the closed system. The closed system is made by two qutrits (A,B) as discussed above. We express the state vector of the additional auxiliary qutrit C as below,
\begin{figure*}
\centering
\includegraphics[scale=1.15]{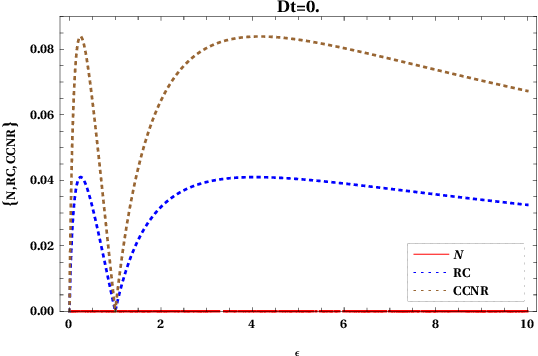}
\caption{Plot of Negativity(N), Realignment(RC) and CCNR vs. $\epsilon$}\label{f1}
\end{figure*}
\begin{equation}
|C\rangle=\alpha|0\rangle+\beta|1\rangle+\gamma|2\rangle\label{eqt}
\end{equation}
with the normalization condition
\begin{equation}
|\alpha|^{2}+|\beta|^{2}+|\gamma|^{2}=1.\label{nc}
\end{equation}
The density matrix of the auxiliary qutrit $|C\rangle$ reads as,
\begin{equation}
\rho_{C}=\left[ \begin{array}{ccc}
\alpha^2&\alpha\beta&\alpha\gamma \\
\alpha\beta&\beta^2&\beta\gamma \\
\alpha\gamma&\beta\gamma&\gamma^2 \\
\end{array}
\right ]. \ \ \ \label{edm'}
\end{equation} 
Using the Eq.\ref{nc} the above density matrix can be written as, 
\begin{equation}
\rho_{C}=\left[ \begin{array}{ccc}
\alpha^2&\alpha\beta&\alpha\sqrt{1-\alpha^2-\beta^2} \\
\alpha\beta&\beta^2&\beta\sqrt{1-\alpha^2-\beta^2} \\
\alpha\sqrt{1-\alpha^2-\beta^2}&\beta\sqrt{1-\alpha^2-\beta^2}&1-\alpha^2-\beta^2 \\
\end{array}
\right ]. \ \ \ \label{edm}
\end{equation}
After interaction the initial density matrix of the open system can be expressed as,
\begin{equation}
\rho_{ABC}(0)=\rho(\epsilon_{1},\epsilon_{2},\epsilon_{3}) \otimes \rho_{C}.  \label{sdm}
\end{equation}

We assume that the auxiliary qutrit C interacts with the qutrit B of the pair through DM interaction. Now the Hamiltonian of the open system can be written as,
\begin{equation}
H=H_{AB}+H_{BC}^{int}. \label{H'}
\end{equation}
Where $H_{AB}$ is the Hamiltonian of qutrit A and qutrit B and $H_{BC}^{int}$ is the interaction Hamiltonian of qutrit B and qutrit C. Here we consider uncoupled qutrit A and qutrit B, so $H_{AB}$ is zero. Now the Hamiltonian of the system becomes,
\begin{equation}
H=H_{BC}^{int}=\vec{D}.(\vec{\sigma_B} \times \vec{\sigma_C}). \label{H''}
\end{equation}
Where $\vec{D}$ is DM interaction strength between qutrit B and auxiliary qutrit C. Here $\vec{\sigma_B}$ and $\vec{\sigma_C}$ denote the Pauli vectors associated to qutrit B and auxiliary qutrit C respectively. We assume that DM interaction exists along the z-direction only. In that  case, the Hamiltonian of the system can be simplified as,
\begin{figure*}
\centering
\includegraphics[scale=1.15]{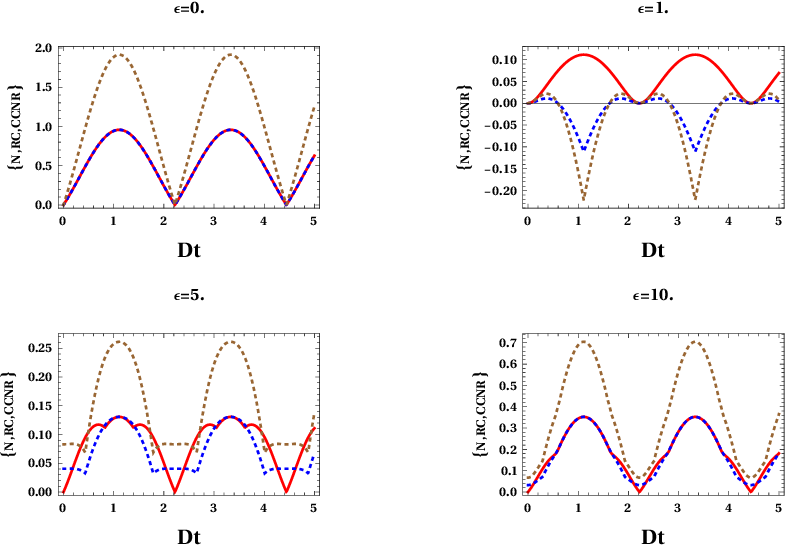} 
\caption{Plot of Negativity(N), Realignment(RC) and CCNR vs. $Dt$}\label{f2}
\end{figure*}
\begin{equation}
H=D.(\sigma_B^X \otimes \sigma_C^Y-\sigma_B^Y \otimes \sigma_C^X). \label{H} \\ 
\end{equation}
Where $\sigma_B^X$, $\sigma_B^Y$ and $\sigma_C^X$, $\sigma_C^Y$ are the X and Y Gell-Mann matrices of qutrit B and qutrit C respectively. The above Hamiltonian is a matrix having $9\times 9$ dimension and is easy to diagonalize by using the method of eigendecomposition. 

\begin{figure}
\centering
\includegraphics[scale=1.15]{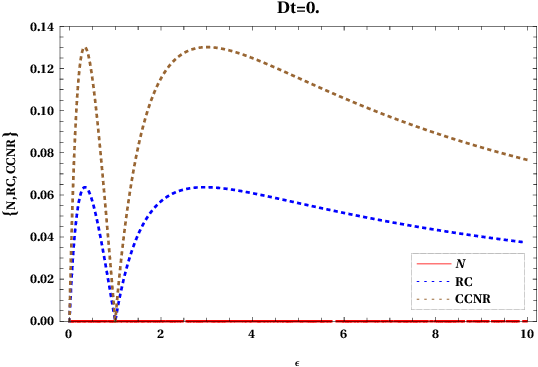}
\caption{Plot of Negativity(N), Realignment(RC) and CCNR vs. $\epsilon$}\label{f3}
\end{figure}
According to the postulate of quantum mechanics, the unitary time evolution of a physical system is obtained from the time-dependent Schr\"odinger  equation given below,
\begin{equation}
i\hbar\frac{d}{dt}|\psi(t)\rangle=E|\psi(t)\rangle.
\end{equation}
Where $E$ is the real energies of the physical system. The solution of the above equation is expressed as,
\begin{equation}
|\psi(t)\rangle=e^{\frac{-iHt}{\hbar}}|\psi(0)\rangle.\label{se}
\end{equation}
For the application of density matrix Eq.\ref{se} can be framed as,
\begin{equation}
\rho(t)=U(t).\rho(0).U(t)^{\dagger}\label{tm1}.
\end{equation}
Where $U(t)=e^{\frac{-iHt}{\hbar}}$ is the unitary matrix, known as `Time Evolution Operator', includes the Hamiltonian $(H)$ in exponential. To simplify the present study we assume $\hbar=1$ and using the Eqs.\ref{sdm} and \ref{tm1} time evolution density matrix of the open system can be written as,
\begin{equation}
\rho_{ABC}(t)=U(t).\rho_{ABC}(0).U(t)^{\dagger}\label{tsdm}.
\end{equation}
This time evolution density matrix is further used to explain the dynamics of the open system.

\section{Negativity, Realignment and CCNR Criteria}
In this section, we discuss about negativity, realignment, and CCNR criteria, used for detection and measurement of entanglement in the paper. The negativity has been used to measure the free entanglement while realignment $(RC)$ and CCNR are both criteria that have been used to detect the bound entanglement of a system. The CCNR criteria have been discovered in two different forms either by cross norms or by realignment of density matrices. This criterion can detect a wide range of bound entangled states over the realignment criterion. In the current work, we have used both criteria to detect the bound entanglement. The negativity $(N)$, realignment $(RC)$ and CCNR criteria are defined as below,

\begin{figure*}
\centering
\includegraphics[scale=1.15]{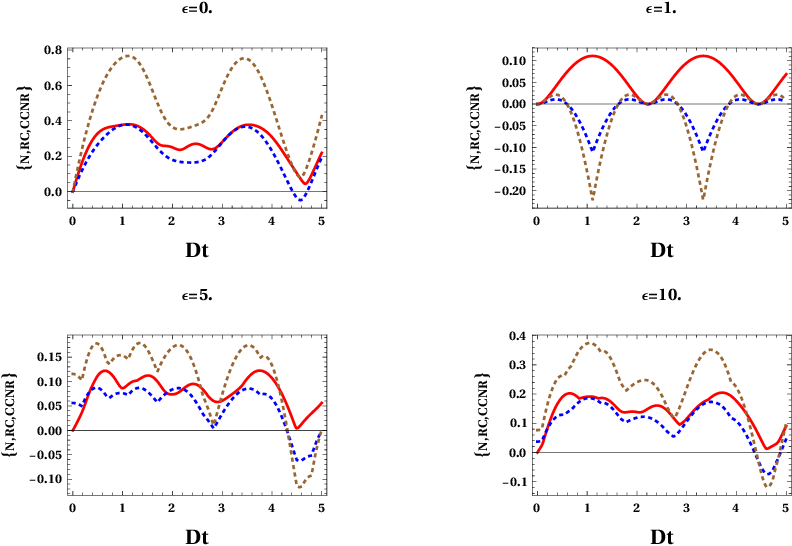} 
\caption{Plot of Negativity(N), Realignment(RC) and CCNR vs. $Dt$}\label{f4}
\end{figure*}
\begin{equation}
N=\frac{(\left \|\rho_{AB}^{T}\right \|-1)}{2}, \label{N}
\end{equation} 
\begin{equation}
RC=\frac{(\left \|\rho_{AB}^{R}\right \|-1)}{2} \label{R}
\end{equation}
and
\begin{equation}
CCNR=\left\|(\rho_{AB}-\rho_{A}\otimes \rho_{B})^R\right\|-\sqrt{(1-Tr \rho_{A}^{2}) (1-Tr \rho_{B}^{2})}. \label{CCNR}
\end{equation}
Where $||..||$, $(..)^T$ and $(..)^R$ represent the trace norm, partial transpose and realignment matrix. Further $\rho_{A}$, $\rho_{B}$ and $\rho_{AB}$ are the reduced density matrices of qutrit A, qutrit B and bound entangled state AB respectively, and expressed as,
\begin{equation}
\rho_{A}=Tr_{BC}(\rho_{ABC}),
\end{equation}
\begin{equation}
\rho_{B}=Tr_{AC}(\rho_{ABC})
\end{equation}
and
\begin{equation}
\rho_{AB}=Tr_{C}(\rho_{ABC}).
\end{equation}

For a system, $N>0$ or $(RC,CCNR)>0$ imply that the state is entangled, $N=0$ and $(RC,CCNR)>0$ imply that the state is bound entangled, and $N>0$ corresponds to the free entangled state.

\section{Open System Dynamics}
In this section, we explore the open system dynamics of the bound entangled state under negativity, realignment, and CCNR criteria using the time evolution density matrix of the system given in Eq.\ref{tsdm}. The dynamical equation of the system involve the parameters $\epsilon_{1},\epsilon_{2},\epsilon_{3}, t$ and DM interaction strength $D$. In the explanation of the current paper, we consider the parameters $D$ and $t$ as a single parameter $Dt$. We know bound entangled state satisfies the trace condition given in Eq.\ref{tr1}. By considering this equation we divided our study into three cases which are given in successive subsections. 

\begin{figure*}
\centering
\includegraphics[scale=1.15]{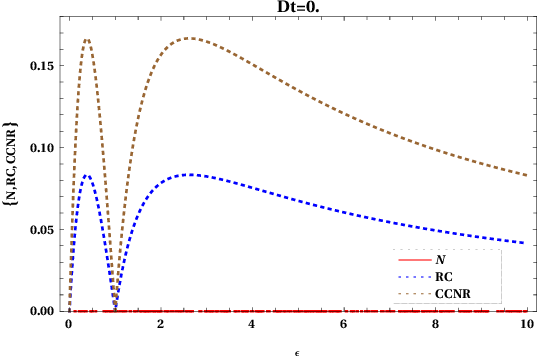}
\caption{Plot of Negativity(N), Realignment(RC) and CCNR vs. $\epsilon$}\label{f5}
\end{figure*}

\subsection*{Case 1: $\epsilon_{1}=\epsilon_{2}=1$, $\epsilon_{3}=\epsilon$}  
In this case, we consider the parameters $\epsilon_{1}=\epsilon_{2}=1$ and explain our study for parameter $\epsilon_{3}=\epsilon$ with the range $0 \leq \epsilon \leq 10$. To proceed the explanation at first we focus on the study of initial condition $(Dt=0)$. Further we enhance the explanation in the range $0 \leq Dt \leq 5$ for different values of $\epsilon$.
In figure \ref{f1} we plot the results for the initial condition, where the solid red line indicates the negativity (N) and dotted blue and brown lines represent the realignment (RC) and CCNR criteria. The figure shows that for $0\leq \epsilon\leq 10$, the negativity is zero $(N=0)$ which implies that initially, the state has no free entanglement. We have noticed that the realignment and CCNR criteria achieve the positive values, except for $\epsilon=1$. Hence the state is separable for $\epsilon=1$, and bound entangled for $0\leq \epsilon< 1$ and $1<\epsilon\leq 10$.

Fig.\ref{f2} displays the dynamical behavior of the bound entangled state with advancement of time $(Dt\geq 0)$ for some values of $\epsilon$.  We have noticed that the increasing strength of DM interaction produces the oscillating free entanglement in the system which also oscillates the realignment and CCNR criteria. Further it is also noticed that the figure obeys the following conditions,
\begin{equation}
(RC,CCNR)>0,N>0,  \label{c1}
\end{equation} 
\begin{equation}
(RC,CCNR)<0,N>0. \label{c2}
\end{equation}
From condition \ref{c1} we can claim that the bound entanglement is converted to free entanglement. On the contrary, the condition \ref{c2} tells that both the criteria failed to detect the bound entanglement in the state but the oscillatory free entanglement is produced in the system due to DM interaction. Here in the second condition, we can not claim the conversion of the bound entanglement to the free entanglement, since we are unable to detect the bound entanglement with this condition. Corresponding to this condition the results can be seen with $\epsilon=1$. Here in the figure, we investigate that the state is completely free from bound entangled to free entangled state with the parameter values $\epsilon=\{0,5,10\}$. 

\subsection*{Case 2: $\epsilon_{1}=1$, $\epsilon_{2}=\epsilon_{3}=\epsilon$}
We explain this case by assuming the parameters $\epsilon_{1}=1$ and $\epsilon_{2}=\epsilon_{3}=\epsilon$, for the range $0\leq \epsilon\leq 10$. The results of this case for initial condition $(Dt=0)$ is shown in fig.\ref{f3}. In this figure, we investigate that for this case under initial condition, the bound entangled state repeats its results as discussed in case one.
\begin{figure*}
\centering
\includegraphics[scale=1.15]{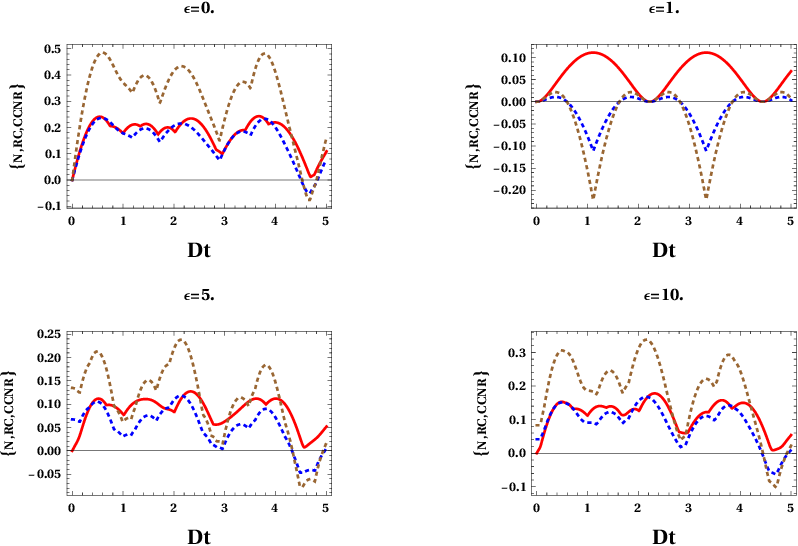} 
\caption{Plot of Negativity(N), Realignment(RC) and CCNR vs. $Dt$}\label{f6}
\end{figure*}

Further, we extend our study towards the dynamical behavior of the bound entanglement with the advancement of time $(Dt>0)$ for some chosen values of $\epsilon$ and corresponding results plotted in fig.\ref{f4}. We have found that in the present case the DM interaction produces the free entanglement with the oscillatory pattern, but it does not exhibit the exact sinusoidal behavior. From the figure it is also observed that the current case follows the conditions (\ref{c1}, \ref{c2}); based on these conditions, one can easily figure out the conversion of bound to free entanglement. 

\subsection*{Case 3: $\epsilon_{1}=\epsilon_{2}=\epsilon_{3}=\epsilon$}
This case starts by considering the parameters $\epsilon_{1}=\epsilon_{2}=\epsilon_{3}=\epsilon$ with the range $0\leq \epsilon\leq 10$. Taking the initial condition $(Dt=0)$ we have displayed the graphical results of the open system in fig.\ref{f5}. Analyzing the results it is noticed that the graphical figures repeat the pattern and show similar results with the previous two cases.

Open system dynamics of the bound entangled state with the advancement of time  $(Dt>0)$ is shown in fig.\ref{f6}. The dynamics are discussed for different values of $\epsilon$ and found that the oscillatory free entanglement is produced in the system due to DM interaction. Although the non-sinusoidal behavior of the entanglement is observed in the system and in the current case the oscillatory behavior is more distorted than the previous one. To study the results shown in the figure, one can recall the conditions (\ref{c1}, \ref{c2}) to get the idea about where the bound entanglement is converted to free entanglement.

\section{conclusion}
In this article, we have studied the open system dynamics of the bound entangled state of two qutrits proposed by Jurkowski \textit{et al}. Under DM interaction we have explored the study in three different cases and found that this interaction produces the free entanglement with negativity measure. In all the cases we also discussed the efficacy of DM interaction to convert the bound entanglement to free entanglement. Here we mention that DM interaction has special qualities as it produces the free entanglement in the system; also it converts the bound entanglement to free entanglement. The bound entangled state dependents on the parameters $\epsilon_{1},\epsilon_{2}$ and $\epsilon_{3}$. We have found the better efficacy of DM interaction with the case one $(\epsilon_{1}=\epsilon_{2}=1,\epsilon_{3}=\epsilon)$ in which the state is completely converted from bound to free entangled state except $\epsilon=1$. On the other hand, the state is always separable for $\epsilon_{1}=\epsilon_{2}=\epsilon_{3}=1$, and for this case, most of the time both the criteria (realignment and CCNR) failed to detect the bound entanglement with the advancement of time $(Dt\geq 0)$, but the free entanglement is produced in the system. In short, corresponding to the state $(\epsilon_{1}=\epsilon_{2}=\epsilon_{3}=1)$ the bound entangled state is fragile in which there is no guarantee of bound to free entanglement conversion. The study can be explored in a broad domain in larger Hilbert spaces to prove the efficacy of DM interaction.

\end{document}